\documentclass{article}

\usepackage{PRIMEarxiv}
\usepackage{amsmath} 
\usepackage[utf8]{inputenc} 
\usepackage[T1]{fontenc}    
\usepackage{hyperref}       
\usepackage{url}            
\usepackage{booktabs}       
\usepackage{amsfonts}       
\usepackage{nicefrac}       
\usepackage{microtype}      
\usepackage{lipsum}
\usepackage{fancyhdr}       
\usepackage{graphicx}       
\graphicspath{{media/}}     

\title{SocialRec: User Activity Based Post Weighted Dynamic Personalized Post Recommendation System in Social Media
}

\author{
  Ismail Hossain, Sai Puppala, Md Jahangir Alam, Sajedul Talukder \\
  School of Computing \\
  Southern Illinois University Carbondale, IL, USA, 62901\\
  \texttt{\{ismail.hossain, saimaniteja.puppala, mdjahangir.alam, sajedul.talukder\}@siu.edu} \\
}
\begin{document}
\maketitle
\begin{abstract}
User activities can influence their subsequent interactions with a post, generating interest in the user. Typically, users interact with posts from friends by commenting and using reaction emojis, reflecting their level of interest on social media such as Facebook, Twitter, and Reddit. Our objective is to analyze user history over time, including their posts and engagement on various topics. Additionally, we take into account the user's profile, seeking connections between their activities and social media platforms. By integrating user history, engagement, and persona, we aim to assess recommendation scores based on relevant item sharing by Hit Rate (HR) and the quality of the ranking system by Normalized Discounted Cumulative Gain (NDCG), where we achieve the highest for NeuMF 0.80 and 0.6 respectively.  Our hybrid approach solves the cold-start problem when there is a new user, for new items cold-start problem will never occur, as we consider the post category values. To improve the performance of the model during cold-start we introduce collaborative filtering by looking for similar users and ranking the users based on the highest similarity scores.

\end{abstract}

\section{Introduction}
\label{sec: introduction}

tent, social media recommendation systems can influence our opinions, shaping public conversations, and impacting our mental health~\cite{feng2024probing}.

Recommendation systems encompass Collaborative Filtering, Content-Based Filtering, and Hybrid Systems. While Content-Based Filtering suggests similar items based on past preferences, it can limit exposure to new content~\cite{cantador2008multi}. Collaborative Filtering, foundational since the 1990s, predicts items aligned with user tastes but faces scalability and sparsity challenges~\cite{barragans2010hybrid}. Recent research focuses on enhancing Collaborative Filtering through new similarity calculations~\cite{liu2014new}, model expansion~\cite{shi2014collaborative}, and user preference augmentation~\cite{kim2010collaborative}.

Hybrid recommendation systems integrate multiple approaches to address issues like the cold start problem~\cite{thakur2022cloud}. However, the current methods often lack a comprehensive consideration of user demographics, history, preferences, and engagement within a single recommendation system.

In our approach, we built a recommendation system for social media that incorporates weighted dynamic demographic data, user post history, and user engagement with posts. This approach aims to solve the cold-start and sparsity problems, leveraging the significant role of demographic attributes like age, gender, and occupation. Without demographic data, a recommendation system would be generic, lacking the personalized touch. Additionally, post categories in social media add a new dimension to user preference and engagement. Introducing demographic attributes as a feature helps overcome the new user problem~\cite{ccano2017hybrid}. However, two users with the same demographic attributes might have different preferences, identifiable through their post history and engagement. User post history refers to the posts on their timeline, while engagement includes reactions, comments, and shares. Considering users' activities within social media, we aim to build a robust recommendation system.

As the first step, we propose a novel approach to calculate the weight of user demographic attributes. Those weights depend on user preferences. For example, a user might prefer sports-related posts over political ones, while another user with the same demographic attributes might prefer science/technology posts over sports. In this scenario, weights are assigned based on different preferences, meaning the weight of demographic data varies due to several user preferences based on post categories. To build a feature set, we use user demographic data, category-wise user post history scores, and engagement scores. Our experiment employs two approaches: matrix factorization and neural network matrix factorization, to observe the results of the recommendation system.

Our study is structured into the following sections. Firstly, in Section~\ref{sec: relatedwork}, we review previous literature on recommendation systems developed across various domains, considering diverse user and item features. Following this, in Section~\ref{sec: methodology}, we provide a comprehensive overview of the architecture of our proposed recommendation system, elucidating the techniques employed. Subsequently, in Section~\ref{sec: experiment}, we present a brief description of the dataset utilized and the results obtained under different scenarios. Section~\ref{sec: application} demonstrates our concept of recommendation systems through a user interface. Finally, we draw our conclusions in Section~\ref{sec: conclusion}.
\begin{figure}
    \centering
    \includegraphics[width=0.7\textwidth]{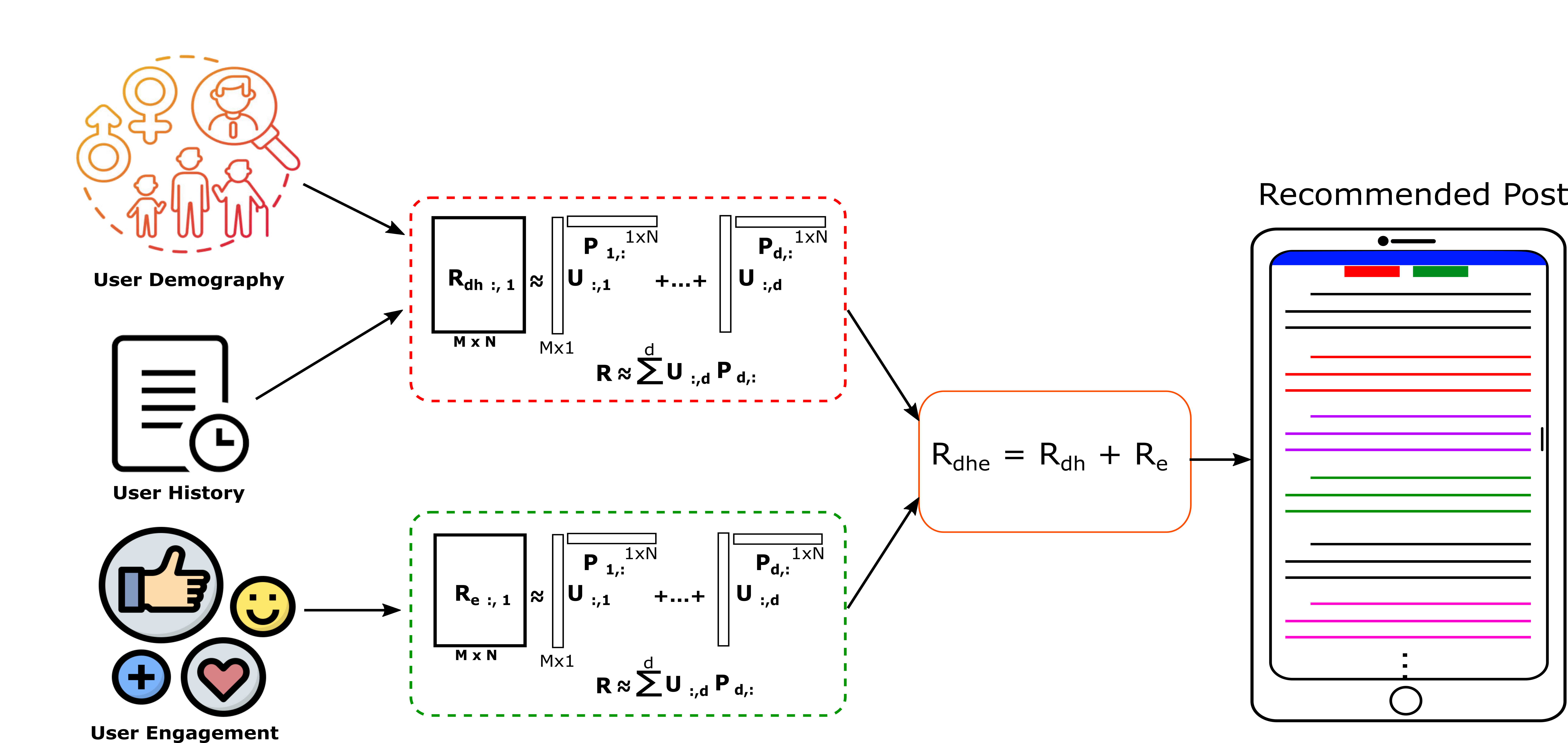}
    \caption{System Architecture for the Recommendation System with three different approaches}
    \label{fig:system-architecture}
\end{figure}

In summary, we introduce the following contributions:

\begin{itemize}
    \item \textbf{Dynamic Calculation of User Demographic Weights:}
    Our approach proposes a method for dynamically calculating weights for demographic attributes based on user posts within social media.
    
   \item \textbf{Incorporating Demographic and Historical Data into Recommendations:}
    We present a recommendation system that integrates user demographic information and historical data.
    
    \item \textbf{Utilizing User Interaction Data for Recommendations:}
    We investigate user interaction data such as reactions, comments, and sharing to develop a recommendation system that incorporates category-wise information from user posts.
    
    \item \textbf{Introducing a Hybrid Recommendation System:}
    We introduce a hybrid recommendation system that combines demographic, historical, and interaction-based data, providing a new dimension to recommendation systems.
\end{itemize}

\section{Related Works}
\label{sec: relatedwork}
Balabanovic and Shoham~\cite{balabanovic1997fab} introduced a system that blends collaborative and content-based recommendation methods to create user profiles. In content-based systems, user preferences are key to generating suggestions, while collaborative approaches identify users with similar tastes and recommendations based on those similarities.

Various methods measure user similarity using preference values, especially explicit ratings like Pearson correlation coefficient (PCC)~\cite{herlocker1999algorithmic, liu2013soco}. Liu et al. utilized PCC and extended it to account for contextual information~\cite{liu2013soco}. Bobadilla et al. proposed JMSD, which blends the Jaccard measure and mean squared difference (MSD) to assess user similarity~\cite{bobadilla2010new}. They considered both the ratio of common ratings and the absolute difference between user ratings.

Liu et al. also introduced the New Heuristic Similarity Model (NHSM), which calculates similarity based on user ratings and global preference in user behavior~\cite{liu2014new}. Zhu et al. weighted popular items in each user-to-item rating matrix and used cosine similarity to calculate user similarity~\cite{zhu2014predict}.

Xu et al.~\cite{xu2013detecting} explored user preferences on microblogs by utilizing information from their connected users. They focused on filtering out unnecessary connections to accurately predict the preferences of specific users, rather than using traditional methods that seek out relevant users.

Servia-Rodriguez et al.~\cite{servia2014tie} incorporated user interactions and social circle information to calculate tie strength between users. They also proposed a personalized model based on tie strength to enhance social services.

Some works, such as those by Wu et al.~\cite{wu2016collaborative} and Li et al.~\cite{li2015deep}, use neural networks like denoising auto-encoders or Restricted Boltzmann Machines (RBM) to model users and items based on the rating matrix. These methods are seen as collaborative approaches because they rely solely on the rating matrix and overlook user demographics, history, and interactions, unlike our method. Elkahky et al.~\cite{elkahky2015multi} developed a multi-view deep model in a joint manner to learn the latent factors of users and items and project them into a shared space. Other approaches, such as those by Pugoy et al.~\cite{pugoy2020bert} and Zheng et al.~\cite{zheng2017joint}, use user reviews to design recommendation systems based on BERT or CNN architectures, respectively.

\section{Problem Statement}
Let $U$ be the set of users and $P$ be the set of items (posts/articles) and demographic data can be defined by $D$, where $D = \{a,g,e,o,l\}$, here, {a: age, g: gender, e: education, o: occupation, l: location, and post categories $C = \{c_1, c_2, c_3, ..., c_k\}$. Each user u is represented by a feature vector $U_f$, which includes demographic information and user history, where $U_f = [\rho_{u,1}, \rho_{u,2}, ...., \rho_{u,d}]$, (here, $ u \in U$, $\rho_{u, i} \in_{1 \leq i \leq 5} D_i$ and $\rho_{u,i} \in_{5 \leq i \leq d} c_k$).
Each item p is represented by a feature vector $P_f$, which captures content-based features (e.g., demographic weights, topic categories), where $P_f = [\mu_{p,1}, \mu_{p,2},...., \mu_{p,d}]$, (here, $ p \in P$, $\mu_{u, i} \in_{1 \leq i \leq 5} D_i$ and $\mu_{u, i} \in_{5 \leq i \leq d} c_k$). Here, $U_f$ and $P_f$ is two 1D vectors having size $d$.

\begin{figure}
    \centering
    \includegraphics[width=0.7\linewidth]{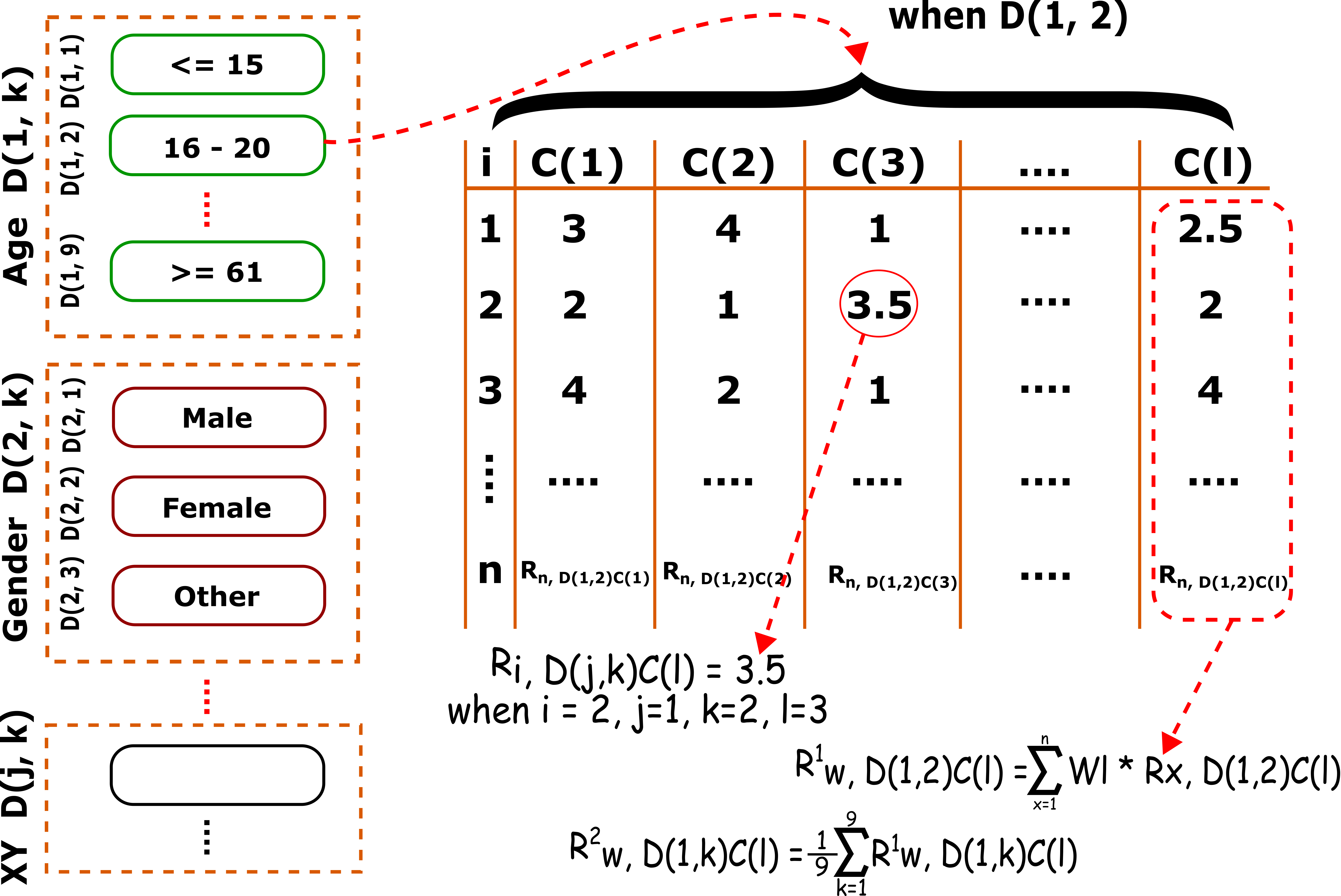}
    \caption{How the weight of each demographic attribute is being calculated under the post category.}
    \label{fig:weight-cal}
\end{figure}

\section{Methodology}
\label{sec: methodology}
\subsection{Weight Calculation}
Let, $R_{i, D(j, k)C(l)}$ is the rating of $i$th participant in the survey who holds a particular demography based on post category.
Here, $D(j,k) \leftarrow$ $k$th type of $j$th demographic attribute.

Suppose, One of the demographic attributes is `Age' and It will be divided into multiple sets of different age ranges.
For example, $\left\{\leq 15, 16-20, 21-26, .... \right\}$. So, $j$th attribute could be `Age' and $k$th type could be `16-20'.

$C(l) \leftarrow$ $l$th category of a post.

If there is total of $n$ participants in the survey then we have X = $\left[R_{1, D(j, k)C(l)}, R_{2, D(j, k)C(l)}, ..., R_{n, D(j, k)C(l)}\right]$.

\[
R^1_{w, D(j, k)C(l)} = \sum^n_{x=1} \omega_{l} \cdot R_{x, D(j, k)C(l)}
\hspace{1cm}
R^2_{w, D(j, k)C(l)} = \frac{1}{K}\sum^K_{k=1} R^1_{w, D(j, k)C(l)}
\]

Here, $R^1_{w, D(j, k)C(l)}$ and $R^2_{w, D(j, k)C(l)}$ are weighted ratings calculated differently and further we consider these as $\omega^1_{dm, l}$ and $\omega^2_{dm, l}$ (weight of $l$th category of a post for $k$th type of $j$th demographic attribute) respectively. $\omega_{l}$ is the weighted average of $l$th category of each post. To calculate $\omega_{l}$, we first determined the median of X.
Let, the median be $R_m$. $\Delta_n = R_m  - R_{n, D(j, k)C(l)}$, and 
$\omega_{l} = \frac{\Delta^{-1}_n}{\sum^n_{x=1} \Delta^{-1}_x}$

Figure~\ref{fig:weight-cal} demonstrates the way of calculating the weights of each demographic attribute. On the left side of the figure shows the all demographic attributes. As an example, we show the dummy data of the user rating table on the right side indicating the survey results against all post categories. Rating is considered on a scale of 0 - 5 and each of them is assigned as $R_{i, D(j, k)C(l)}$ which is further used to calculate the weight.

\subsection{Matrix Factorization}
Let us have a total of $n$ users in one friends or follower list. Meanwhile, we have a total of $m$ items (articles or posts). We define the user-post interaction matrix $R_o^{n \times m}$ from users' implicit data as,

\[
r_{o, up} = 
\begin{cases}
    1 & \text{if interaction (user } u \text{, post } p \text{) is observed;} \\
    0 & \text{otherwise}
\end{cases}
\]

Here a value of 1 for $r_{o, up}$ indicates that there is an interaction between user $u$ and post $p$; Interaction means $u$ could post $p$ something, or put reaction on $p$ or comment on $p$ or can share $p$. Similarly, a value of 0 means the user doesn't have any interaction with the post $p$, it can be that the user is not aware of the post. This poses challenges in learning from implicit
data since it provides only noisy signals about users’ preferences. While observed entries at least reflect users’ interest in posts, the unobserved entries can be just missing data and there is a natural scarcity of negative data.

\textbf{User Demography and History Based.} 
Now we have user feature embedding $U_1 \in \mathbb{R}^{n \times d}$ and item feature embedding $P_1 \in \mathbb{R}^{m \times d}$.

Let, 
\begin{equation}
U_1 = \begin{bmatrix}
U_{1,f} \\
U_{2,f} \\
\vdots \\
U_{n,f}\\
\end{bmatrix}
P_1 = \begin{bmatrix}
P_{1,f} \\
P_{2,f} \\
\vdots \\
P_{m,f}
\end{bmatrix}
\end{equation}

Here, $U_{n,f} \in U_1$ we have $\rho_{n,d} \in U_{n,f}, n \in U$ and, $I_{m,f} \in P_1$ we have $\mu_{m,d} \in I_{m,f}, m \in P$, so, we calculate $\rho_{n, d}$ and $\mu_{m,d}$ in following way.

\[
\rho_{n, d} = \left\{ 
\begin{array}{rcl}
\omega^1_{\text{avg}, i} & \text{if $i$th demographic attribute} \\ 
c_i & \text{if $i$th post category}
\end{array}
\right.
\]

\[
\mu_{n, d} = \left\{ 
\begin{array}{rcl}
\omega^2_{\text{avg}, i} & \text{if $i$th demographic attribute} \\ 
\Tilde{c}_i & \text{if $i$th post category}
\end{array}
\right.
\]

\[
\omega^1_{\text{avg}, i} = \sum_{j=1}^k X_j \cdot \omega^1_{dm, j}
\hspace{1cm}
\omega^2_{\text{avg}, i} = \sum_{j=1}^k X_j \cdot \omega^2_{dm, j}
\]

\[
X_j \in \left[ \frac{\delta^{-1}_1}{\sum_{i=1}^k \delta^{-1}_i}, \frac{\delta^{-1}_2}{\sum_{i=1}^k \delta^{-1}_i}, \dots, \frac{\delta^{-1}_k}{\sum_{i=1}^k \delta^{-1}_i} \right]
\hspace{1cm}
\delta_j = \eta (\omega^1_{dm, j} \cdot c_j) + (1 - \eta) (\omega^2_{dm, j} \cdot \Tilde{c}_j)
\]

Here, $\omega^1_{dm, j}$ indicates the weight of the $j$th category of the post history under a certain type of demographic attribute for user latent factors and $\omega^2_{dm, j}$ indicates the weight of the jth category of each post under a certain type of demographic attribute for post latent factors. Here, $c_j$ indicates the probability scores of the $j$th category of the user's entire post history and $\Tilde{c}_j$ indicates the probability score of the jth category of a certain post. $\eta$ is a control parameter having a boolean value, so, when $\eta = 1$, $\delta_j$ will be calculated for user feature embedding $U_1$ and for $\eta = 0$, item feature embedding $P_1$.

The matrix factorization result is the user-post interaction matrix $R_{dh}$ as the product of two lower-dimensional matrices: a user embedding $U_1$ and an item embedding $P_1$.
\[
R_{dh} \approx U_1 \odot P_1^T
\]
Here, the product $U_1P_1^T$ gives the predicted recommended scores for all user-post pairs.

In collaborative filtering, we start with a user-post interaction matrix $R_{dh}$, where rows represent users, columns represent items, and the entries represent ratings or interactions between users and items.
\[
R_{dh} = \begin{bmatrix}
r_{1,1} & r_{1,2} & \dots & r_{1,m} \\
r_{2,1} & r_{2,2} & \dots & r_{2,m} \\
\vdots & \vdots & \ddots & \vdots \\
r_{n,1} & r_{n,2} & \dots & r_{n,m}
\end{bmatrix}
\]
Here, $r_{i,j} = \textstyle f (i, j | U_{1}, P_{1}) = U_{1,i} \odot P_{1, j}^T = \sum^d_{k=1}u_{ik}p_{jk}$ represents the rating given by user $i$ to item $j$.
Now we have a user-post matrix and it belongs to interaction vector $[r_{n,1}, r_{n,2}, r_{n,3}, ..., r_{n,m}]$ for $n$th user with $m$ items. This vector gives us the recommendation scores for all posts within his friends circulated throughout the social media. We can recommend top K posts to the user.

\textbf{User Engagement Based.} Let $U_2 \in \mathbb{R}^{n \times d}$ be the user feature matrix and $P_2 \in \mathbb{R}^{m \times d}$ be the item (article/post) feature matrix. Here, n represents the total number of users (friends or followers including oneself), m is the total number of items (articles/posts) circulated within that user timeline in the social media, and d is the total number of features $U_2$ and $P_2$ contain individually.
Let,
\begin{equation}
U_2 = \begin{bmatrix}
\phi_{1,1} & \phi_{1,2} & \dots & \phi_{1,d} \\
\phi_{2,1} & \phi_{2,2} & \dots & \phi_{2,d} \\
\vdots & \vdots & \ddots & \vdots \\
\phi_{n,1} & \phi_{n,2} & \dots & \phi_{n,d}
\end{bmatrix}
\hspace{1cm}
I_2 = \begin{bmatrix}
\omega_{1,1} & \omega_{1,2} & \dots & \omega_{1,d} \\
\omega_{2,1} & \omega_{2,2} & \dots & \omega_{2,d} \\
\vdots & \vdots & \ddots & \vdots \\
\omega_{m,1} & \omega_{m,2} & \dots & \omega_{m,d}
\end{bmatrix}
\end{equation}

Here, 
\[
\phi_{i,j} = \frac{(\sum^R_r  w_r * \eta_r) + w_c * \eta_c + w_s * \eta_s}{\sum \eta_{r} + \eta_c + \eta_s}
\]
\[
\omega_{i,j} = c_{ij}
\]

We have $\eta_{r}$ which indicates the count of a certain reaction $r \in \Tilde{R}$ of $i$th user put on posts for the $j$th category. Current social medias use reactions- $\Tilde{R}=$[ like, haha, love, angry, care, sad, ... ]. Here, $w_r$ denotes the weight of reaction $r$ among $\Tilde{R}$. $w_c$ and $w_s$ are weights for commenting and sharing respectively. And, $\eta_c$ and $\eta_s$ denote the number of comments and shares respectively for the $j$th category of the $i$th user.
$c_{ij}$ indicates the probability value of the $j$th category of $i$th post.

Now, if we perform a dot product between $U_2$ and $P_2$
it results in a user-post interaction matrix or utility matrix $R_e$.

\[
R_{e} \approx U_2 \odot P_2^T
\]
\textbf{Hybrid Approach:}
In recent years, several matrix factorization models have been created to take advantage of the growing amount and diversity of available interaction data and use cases. Hybrid matrix factorization algorithms can combine both explicit and implicit interactions~\cite{zhao2016hybrid}, as well as merge content-based and collaborative data~\cite{zhou2012kernelized,adams2010incorporating,fang2011matrix}.

If a user hasn't posted anything since his account opening, he will not have any post history but might have an engagement to other posts. So, to solve this problem we can aggregate $R_{dh}$ and $R_{e}$ that is shown in Figure~\ref{fig:system-architecture}.

\begin{equation}
    R_{dhe} \approx R_{dh} + R_{e}
\end{equation}
    
So, when there is a new user and new item that could create a `cold-start' problem~\cite{schein2002methods, BOBADILLA2012225, LIKA20142065}. This problem can be solved using several techniques where there is a new user by ~\cite{ schein2002methods, BOBADILLA2012225, feng2021rbpr, sahebi2011community, herce2020new, tahmasebi2021hybrid}, and when there is a new item by~\cite{wei2021contrastive, natarajan2020resolving} and new user and new item both are solved by~\cite{LIKA20142065}.
So, our equation (1) will resolve the cold start when there is a new item or new user who has post engagement but will fail to solve for a new user who doesn't have any social activities (post history and post engagement both).

This problem can be solved based on the similarity of user demographic data (e.g. age, gender, and occupation). We consider demographic similarity between the users to find better neighbors for them. The demographic similarity is based on the weighted average of demographic attributes such as age, gender, and occupation which can be calculated as follows~\cite{tahmasebi2021hybrid}:

\begin{equation}
Sim (u, v)_{dm}  = \frac{\sum_{j=1}^{\alpha} s_j * \omega_{dm}^{(j)}}{\sum_{j=1}^{\alpha} \omega_{dm}^{(j)}}
\end{equation}

where \[Sim (u, v)_{dm} \in [0, 1] \] is the demographic similarity between the users $u$ and $v$, $s_j$ is the
similarity value of the jth attribute of the users $u$ and $v , \omega_{dm}^{(j)}$ is the weight of jth attribute out of $\alpha$ attributes, and l is the number of all demographic attributes. It should be noted that if the jth attribute of two users is equal then the value of the similarity $s_j$ will be 1, otherwise the value will be 0.

Suppose, we have a total of $n$ users set $V = \{v_1, v_2, ..., v_n\}$ who have social media activities, by applying Equation (2) we can calculate the similarity of a new user $u$ and $v_i$, $v_i$ $\in V$ and then will have similarity matrix $S$.
\[
S (u | v_1, v_2,....v_n) = \begin{bmatrix}
s_1 \\
s_2 \\
\vdots \\
s_n\\
\end{bmatrix}
\]

The top \( k \) similarity scores of \( S \) can be represented as:
\[
\begin{array}{ccccccc}
\top & s_{i, (1)} & \top & s_{j, (2)} & \top & \ldots & \top \\
\end{array}
\]

Here, \( s_{i, (1)} \) and \( s_{j, (2)} \) represent the first largest similarity between $u$ and $v_i$ and second largest similarity between $u$ and $v_j$ from \( S \) respectively, where i and j are the two different indices of $V$. Now, we have a similarity matrix with top \( k \) similarity scores and \( R_{dhe} \) from Equation (1).
We can calculate the recommendation scores of \( m \) items for the new user \( u \).

Let, 
\begin{equation}
S_{t} = \begin{bmatrix}
    s_{i (1)}\\
    s_{j, (2)}\\
    \vdots \\
    s_{x, (k)}
\end{bmatrix}
M_{t} = \begin{bmatrix}
    M_{dhe, i}\\
    M_{dhe, j}\\
    \vdots \\
    M_{dhe, x}
\end{bmatrix}
\end{equation}

\begin{equation}
    R_{u} \approx S_{t}^T M_{t}
\end{equation}

Here, $ R_{u} \in \mathbb{R}^{1 \times m} $ is the recommendation matrix with scores for $m$ posts to the new user $u$. 

\begin{figure}
    \centering
    \includegraphics[width=0.7\textwidth]{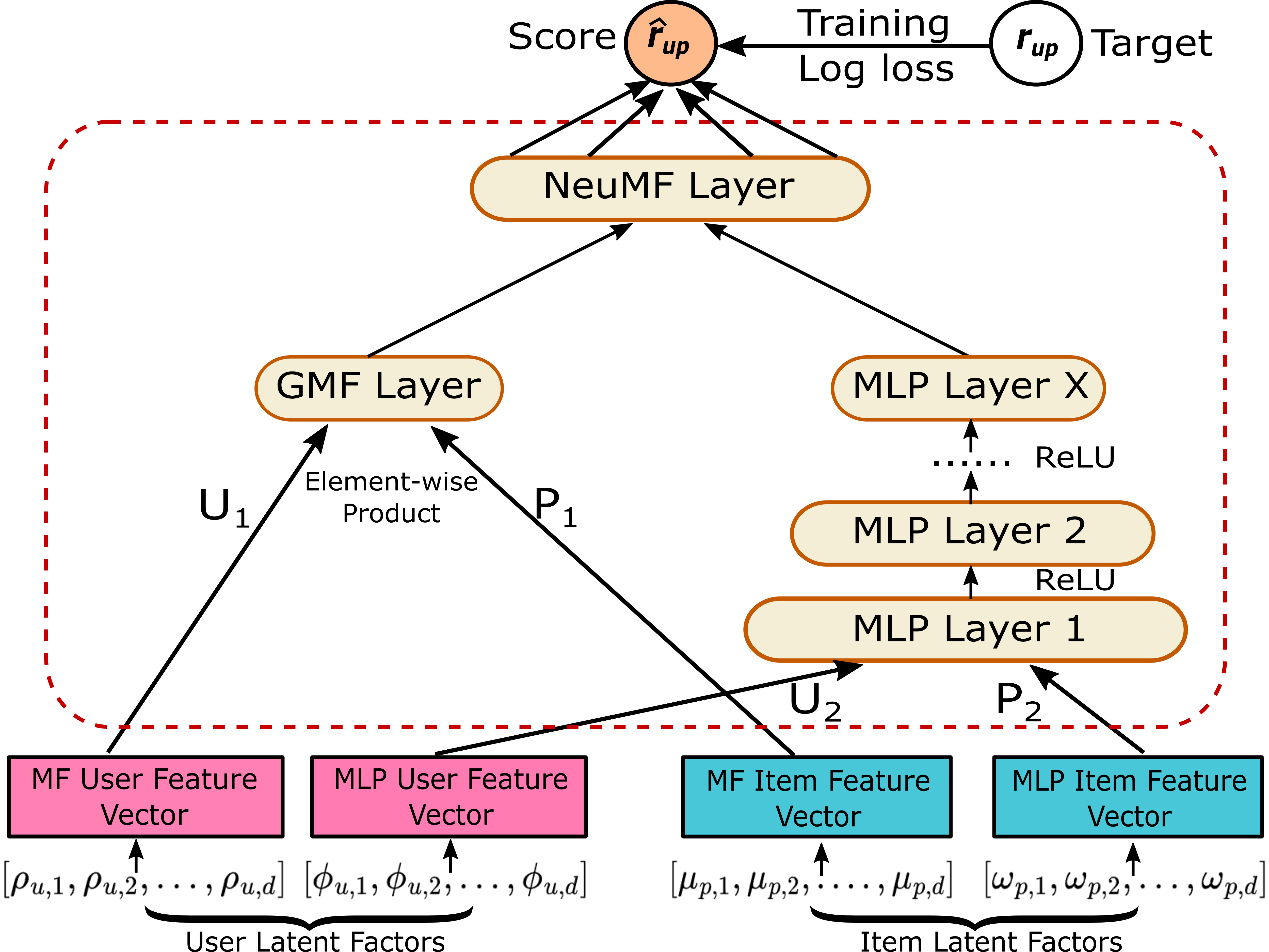}
    \caption{Architecture of Neural Matrix Factorization}
    \label{fig:mcf-archi}
\end{figure}

\subsection{Neural Matrix Factorization}
\textbf{GMF.} We have been utilizing matrix factorization (MF) with a straightforward and fixed inner product to estimate user-item interactions. However, this approach has limitations when it comes to capturing complex user-item interactions~\cite{he2017neural}. As MF is a popular recommendation model that has been widely researched in the literature, incorporating it enables neural collaborative filtering (NCF) to replicate a broad range of factorization models~\cite{rendle2010factorization}. The NCF framework allows the generalization and expansion of MF. Employing non-linear activation functions can modify MF into a non-linear setting, which may prove to be more expressive than the linear MF model~\cite{he2017neural}.

\textbf{MLP.} Typically, neural collaborative filtering (NCF) uses two pathways to model users and items. Consequently, it seems natural to merge the features from the two pathways through concatenation. This approach has been widely employed in multimodal deep learning research~\cite{srivastava2012multimodal, zhang2014start}. However, a simple concatenation of vectors does not capture interactions between the user and item latent features, which falls short in modeling the collaborative filtering effect. To address this, we propose integrating hidden layers on the concatenated vector, leveraging a standard multilayer perceptron (MLP) to learn the interaction between user and item latent features. This approach provides the model with greater flexibility and non-linearity, enabling it to learn the interactions between $U_2$ and $P_2$ more effectively, unlike the generalized matrix factorization (GMF) approach, which relies solely on a fixed element-wise product.

Generalized Matrix Factorization (GMF) uses a linear kernel to model latent feature interactions, while the Multilayer Perceptron (MLP) employs a non-linear kernel to learn the interaction function from data. By fusing GMF and MLP, we can enhance their ability to model complex user-item interactions by allowing them to reinforce each other.

To achieve this, GMF and MLP share the same embedding layer and then combine the outputs of their interaction functions. This approach aligns with the concept of the well-known Neural Tensor Network (NTN)~\cite{socher2013reasoning}. The combination of GMF with a one-layer MLP can be formulated as follows:

\[ \hat{r}_{up} = \sigma ( h^T a (U_1 \odot P_1 + W \left[ \begin{array}{c}
U_2 \\ P_2 \end{array} \right] + b)) \]

However, sharing embeddings between GMF and MLP may limit the performance of the fused model. For instance, this approach necessitates that both GMF and MLP use the same size of embeddings; for datasets where the optimal embedding sizes for the two models differ significantly, this solution might not yield the optimal ensemble.

To enhance the flexibility of the fused model, we allow GMF and MLP to learn separate embeddings and combine the two models by concatenating their final hidden layers. This approach is illustrated in Figure~\ref{fig:mcf-archi}, and its formulation is provided below:
\[
\Phi^{GMF} = U_1 \odot P_1
\]
\[
\Phi^{MLP} = aL (W^T_L (a_{L-1}(...a_2(W^T_2 \left[\begin{array}{c}
    U_2  \\
    P_2
\end{array} \right] + b_2)...)) + b_L)
\]
\[
\hat{r}_{up} = \sigma(h^T \left[\begin{array}{c}
     \Phi^{GMF} \\
     \Phi^{MLP}
\end{array} \right])
\]

\begin{figure}
\centering
\includegraphics[width=\textwidth]{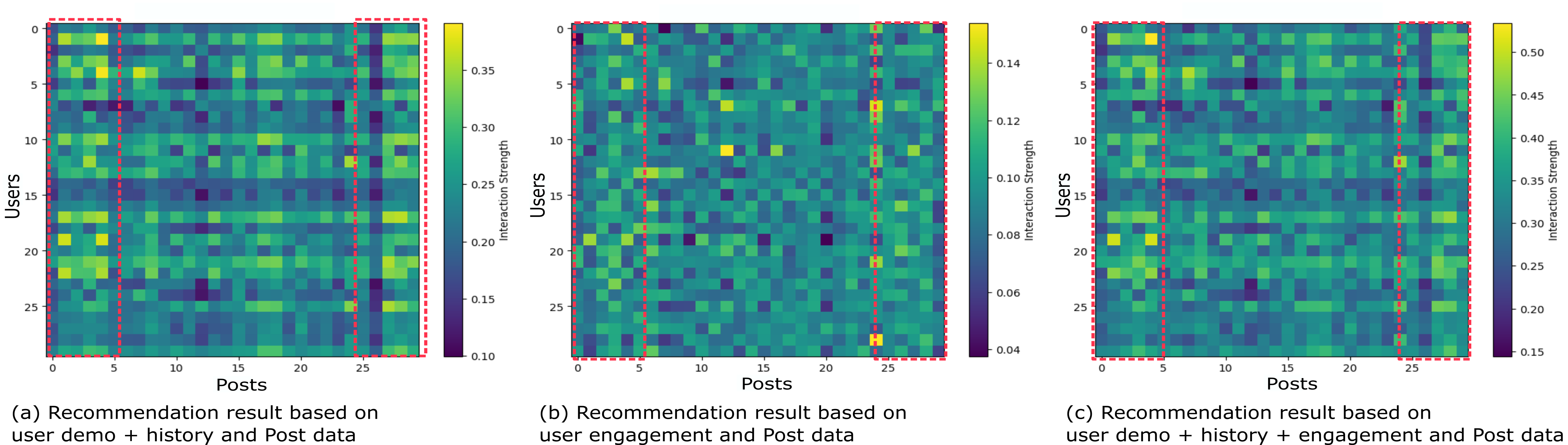}
\caption{A demonstration of a recommendation system for three different scenarios by considering (a) users' demographic data and History (b) user engagements or interactions, and (c) result of Hybrid approach}
\label{fig:experiment-1}
\end{figure}

\section{Experiment}
\label{sec: experiment}

\subsection{Dataset}
\label{subsec:dataset}
Sometimes, domain-specific datasets are not readily available or open to the public, making it difficult to collect the data needed for research purposes ~\cite{guo2024generative}. In our case, collecting demographic data, user history, and engagement data is challenging due to data extraction restrictions and associated costs on many social media platforms. To continue our study, we experimented with a synthetic dataset. We required three datasets to represent user social media activities: user demographics, user post history, and user post engagement or interaction.

We considered demographic attributes such as age, gender, occupation, education, and location. User history and engagement were categorized into ten different areas: `science', `technology', `entertainment', `sports', `finance', `art', `education', `travel', `health', and `politics'. These categories are predominant, with social media posts related to them circulating frequently. Additionally, we included user interactions with posts, such as reactions, sharing, and commenting. Each user was assigned a random number of posts (0 to 10) out of 2000, with 500 users. In our synthetic dataset, we recorded the percentage of user history and interaction for each category.

Instead of using direct demographic values to generate recommendation scores, we weighted each demographic attribute. As previously mentioned, each attribute significantly influences user activities on social media. We experimented with different combinations of weights within the range of [0.1, 0.6] for the features(e.g., age, gender, occupation) as described in ~\cite{tahmasebi2021hybrid}, to observe their impact on a hybrid recommendation system. Instead of defining constant values as weights for each attribute, we relied on survey data. 

We conducted a survey, distributing a questionnaire to analyze user data and determine the weight of each demographic attribute. The questionnaire included questions: (i) What are the user's preferences by post category? (ii) How do demographic attributes influence the likelihood of a post related to sports, technology, health, politics, etc.? 

Our synthetic dataset includes user demographics, user history for each post category, and user engagement/interaction for each post category. It provides the percentage value for each category, indicating how often users post or interact (react, share, comment) with posts over time.

\subsection{Result}
\textbf{MF without Cold Start:} In our MF approach, we analyze user post-interaction through a user-feature matrix multiplied by a post-feature matrix. Figure~\ref{fig:experiment-1} presents three scenarios: (a) incorporating demographic and historical data, (b) focusing on user engagement and post categories, and (c) combining both. Variations in recommendation scores highlight the influence of feature selection.

\textbf{MF with Cold Start:} Figure~\ref{fig:experiment-2a} shows experiments with new users lacking interaction history. Recommendation scores for them are initially zero but improve when demographic, history, and interaction data are combined (Figure~\ref{fig:experiment-2b}(a)). Comparing new users to similar existing ones significantly enhances results, effectively addressing the cold start issue (Figure~\ref{fig:experiment-2b}(b)). New post introductions don't trigger cold start issues due to categorical probability scores based on content.

\begin{figure}
\centering
\includegraphics[width=0.8\textwidth]{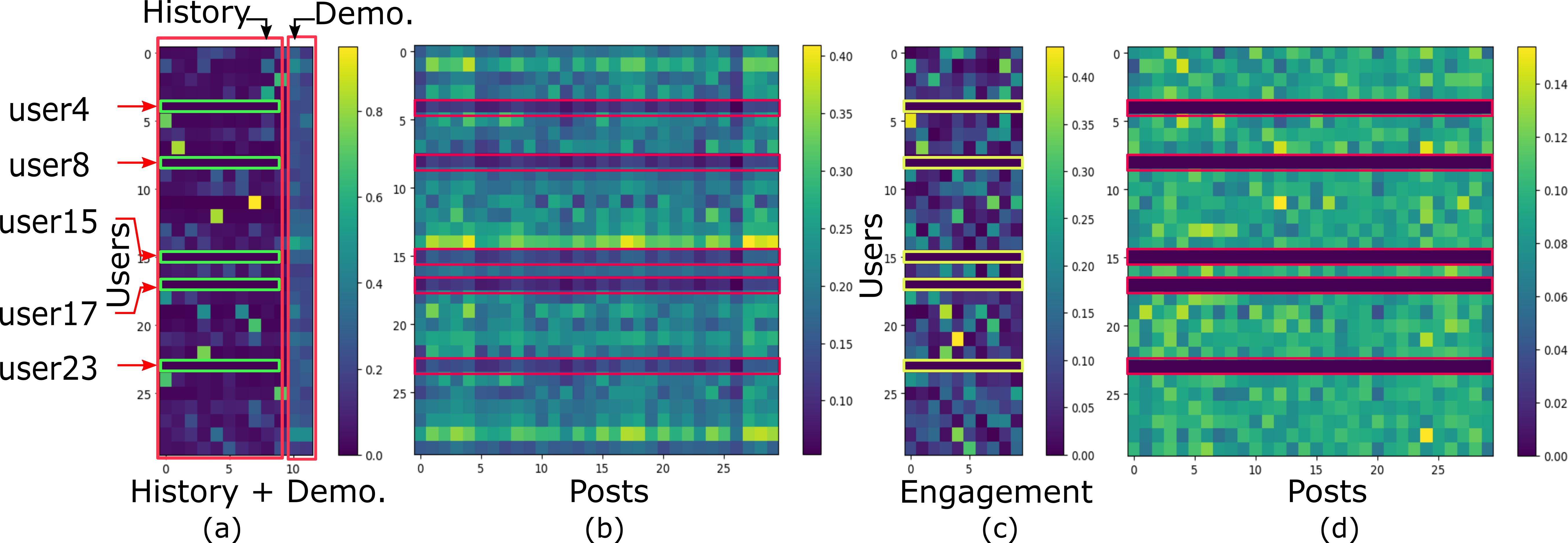}
\caption{Cold start scenario is shown at (a) User feature matrix, (b) Factorization result of User and Post when History and Demographic data are considered as features set, (c) User feature matrix based on engagement, (d) Factorization result of User and Post based on engagement. }
\label{fig:experiment-2a}
\end{figure}

\begin{figure}
\centering
\includegraphics[width=0.8\textwidth]{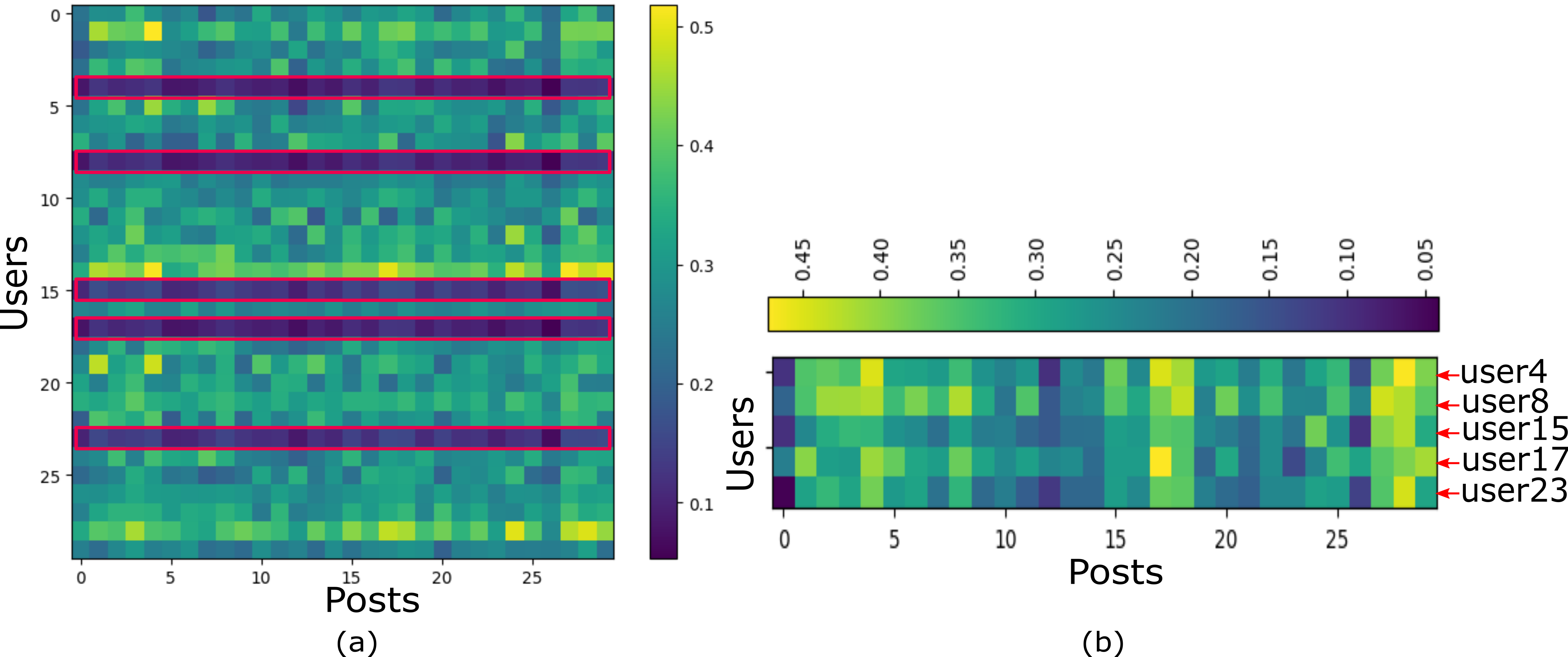}
\caption{The output of hybrid system when the recommendation is not strong at (a) and Collaborative Filter based similarity-based recommendation result at (b)}
\label{fig:experiment-2b}
\end{figure}

\textbf{NeuMF.}
We conducted experiments using neural matrix factorization models, specifically exploring our NeuMF model, which combines the GMF and MLP models. These three models were fine-tuned with our dataset. For GMF, we incorporated user latent factors and post latent factors, which encompass user demographic attributes and post categories. Meanwhile, for MLP, we utilized latent factors derived from user engagement data and post categories. NeuMF was fine-tuned by leveraging pre-trained GMF and MLP models. Figure~\ref{fig:experiment-3} illustrates the performance of these three models, showcasing metrics such as Loss, Hit Rate (HR) (to understand the rate of relevant item sharing to the user), and Normalized Discounted Cumulative Gain (NDCG) (to understand the quality of the ranking system in terms of relevant position) with 10 recommended posts. Our model was trained for 73 epochs with learning rate 1e-03, batch size 128, and three hidden layers within multi-layer perceptons~\cite{he2017neural}, and NeuMF consistently outperformed GMF and MLP across these metrics (Loss, HR, NDCG).

\begin{figure}
\centering
\includegraphics[width=\textwidth]{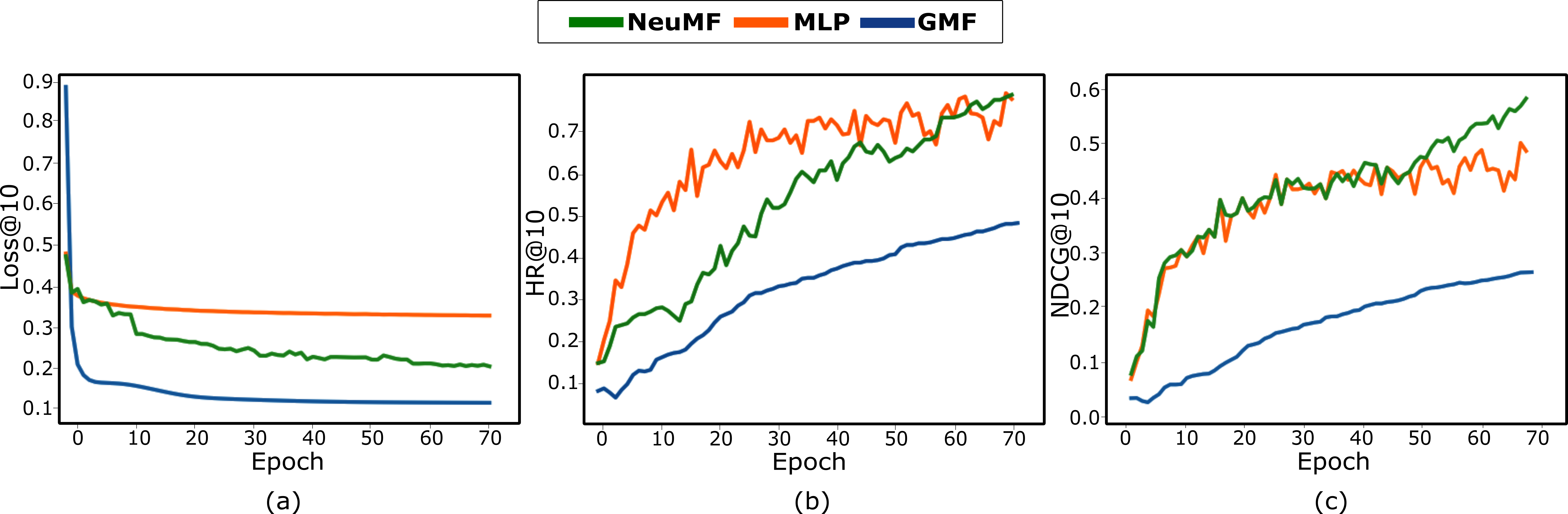}
\caption{Performance of three models with respect to (a) Loss, (b) Hit Rate (HR), and (c) Normalized Discounted Cumulative Gain (NDCG)}
\label{fig:experiment-3}
\end{figure}

\subsection{Discussion}
Recommender systems encounter several challenges that can impact their effectiveness, making it essential to develop and implement effective solutions.
Our approach addresses familiarity bias by leveraging user demographics, past interactions, and engagement. To tackle the "cold start" problem, we use post categorization, allowing new users to contribute demographic information. By organizing user history by demographics and categories, we mitigate sparsity and provide valuable data. Despite the computational demands of Matrix Factorization, our system achieves accuracy by incorporating user demographics and implicit feedback. Additionally, category scores enhance recommendation diversity across different categories.
\begin{figure}
    \centering
    \includegraphics[width=0.8\textwidth]{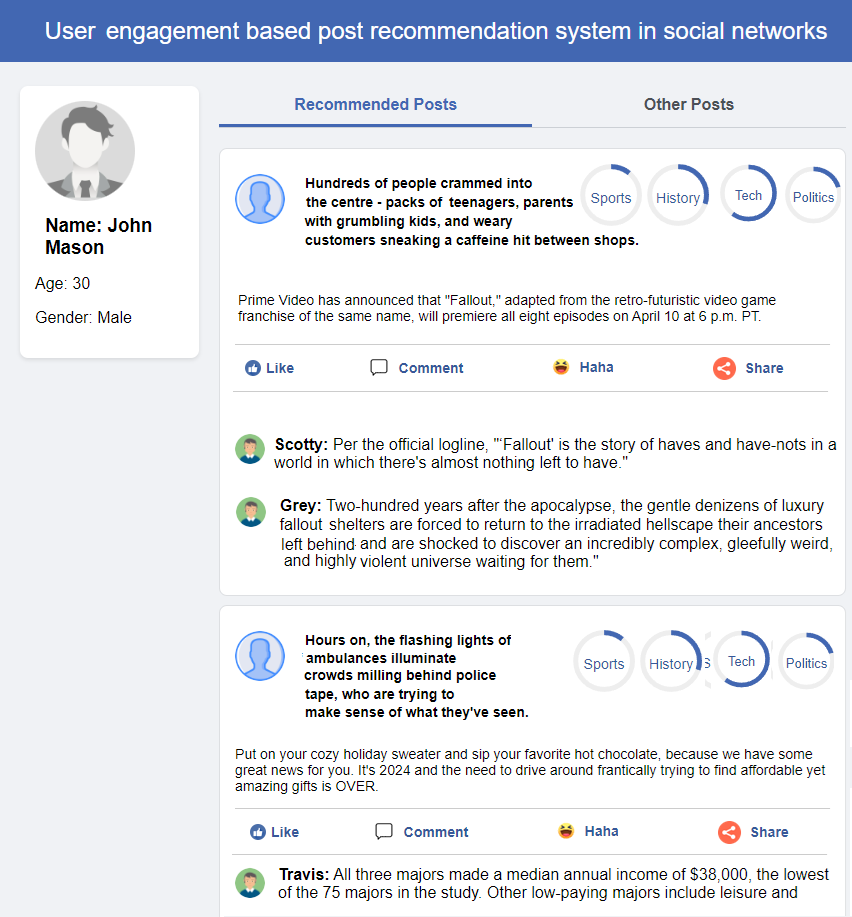}
    \caption{User Interface to show the output by deploying the model in the social media.}
    \label{fig:application}
\end{figure}
For reliability, we prioritize recommending posts circulating among a user's followers or friends, especially if the user hasn't seen them yet. However, considering user demographics and preferences can impact privacy.
Although we used a synthetic dataset for this study, similar data could be obtained from real-life sources. Synthetic data has concerns like 'hallucination' and 'correctness and diversity'~\cite{guo2024generative}. To ensure relevant posts are shared, we focus on common topics and use multi-level classification to generate probability scores for each category.
User habits can change due to age, education, and social influences, so our recommendation system is dynamic. However, we have not yet incorporated changes in user history or interactions over time, which could further enhance our approach.

Lastly, our experiments with NeuMF have been computationally intensive.

\section{Application}
\label{sec: application}
we have shown a user interface in Figure~\ref{fig:application} to demonstrate our idea of RS. By showing the post categories with the circular progression that indicates how much a post falls under each category. We tried to build a UI similar to a few social medias to make it clear to the others. In our system, posts will be recommended based on their ranks, and the scores of each post's category are shown next to the user post. This system can be used for two different purposes (1) the user can see only the recommended posts (2) a user has the chance to see the other posts.

\section{Conclusion}
\label{sec: conclusion}
We have developed a recommendation system for social medias that takes into account demographic attributes, user post history, and user engagement. By categorizing user posts, we use a dynamic weight calculation approach to create a more robust recommendation system. Both user post history and user engagement play key roles in shaping the system based on user preferences.
Rather than assigning a single category to a post, we consider the probabilities of different categories for each post. This allows for multiple preferences and varying category scores, which in turn adjust the weight of the demographic data for each user.
We presented two approaches: traditional Matrix Factorization and a Neural Network model. We conducted experiments to evaluate model performance using our feature set in these two different setups. Additionally, we proposed methods to address the cold-start problem for both new users and new items. 


\end{document}